\documentclass[runningheads]{llncs}
\usepackage[T1]{fontenc}
\usepackage{cite}
\usepackage{times}
\usepackage{amsmath,amssymb,amsfonts}
\usepackage{algorithm,algorithmic}
\usepackage{subfigure}
\usepackage[pdftex]{graphicx}
\usepackage{textcomp}
\usepackage{color}
\usepackage{xcolor}
\usepackage{pgfplots}
\pgfplotsset{compat=1.16}
\usepackage{tikz}
\usepackage{bm}
\usepackage[pagebackref=true]{hyperref} 
\usepackage{color,soulutf8}
\usepackage{booktabs}
\usepackage{multirow}
\usepackage{amsmath}
\usepackage{amsfonts,bm}
\usepackage{graphicx,lipsum}
\usepackage{todonotes}
\usepackage{floatpag}

\newboolean{useComments}
\setboolean{useComments}{FALSE}
\definecolor{dark_purple}{rgb}{0.1, 0.0, 0.4}
\definecolor{dark_green}{rgb}{0.0,0.2,0.5}
\definecolor{dark_red}{rgb}{0.85,0, 0}
\ifthenelse{\boolean{useComments}}{%
	\newcommand{\mh}[1]{\todo[inline,color=white!40,bordercolor=white]{\textcolor{teal!70!black!80}{\textbf{Moritz:}\textmd{\;#1}}}} 
	\newcommand{\ad}[1]{\todo[inline,color=white!40,bordercolor=white]{\textcolor{blue}{\textbf{Akshay:}\textmd{\;#1}}}}
	\newcommand{\vh}[1]{\todo[inline,color=white!40,bordercolor=white]{\textcolor{purple}{\textbf{Vahid:}\textmd{\;#1}}}}
}{%
	\newcommand{\mh}[1]{}
	\newcommand{\ad}[1]{}
	\newcommand{\vh}[1]{}
}

\newcommand{\etal}{\MakeLowercase{\textit{et al.}}}

\begin{document}
\title{Backdoor Mitigation in Deep Neural Networks via Strategic Retraining \thanks{This research was funded in part by
    the EU under project 864075 CAESAR,
    the project	Audi Verifiable AI,
    and the BMWi funded KARLI project (grant 19A21031C).}}
\titlerunning{Backdoor Mitigation in Deep Neural Networks via Strategic Retraining}
\author{Akshay Dhonthi\inst{1,2} \and
Ernst Moritz Hahn\inst{2} \and
Vahid Hashemi\inst{1}}
\authorrunning{A. Dhonthi et al.}
\institute{AUDI AG, Auto-Union-Stra\ss e 1, 85057, Ingolstadt, Germany \and
Formal Methods and Tools, University of Twente, Enschede, Netherlands\\}
\maketitle              % typeset the header of the contribution

\begin{abstract}
    Deep Neural Networks (DNN) are becoming increasingly more important in assisted and automated driving.
    Using such entities which are obtained using machine learning is inevitable:
    tasks such as recognizing traffic signs cannot be developed reasonably using traditional software development methods.
    DNN however do have the problem that they are mostly black boxes and therefore hard to understand and debug.
    One particular problem is that they are prone to hidden \emph{backdoors}.
    This means that the DNN misclassifies its input, because it considers properties that should not be decisive for the output.
    Backdoors may either be introduced by malicious attackers or by inappropriate training.
    In any case, detecting and removing them is important in the automotive area, as they might lead to safety violations with potentially severe consequences.
    In this paper, we introduce a novel method to remove backdoors.
    Our method works for both intentional as well as unintentional backdoors.
    We also do not require prior knowledge about the shape or distribution of backdoors.
    Experimental evidence shows that our method performs well on several medium-sized examples.
  \end{abstract}

  \section{Introduction}
Advanced Driver Assistive System (ADAS) or Autonomous Driving (AD) functions \cite{fingscheidt2022deep} generally use Deep Neural Networks (DNN) in their architecture to perform complex tasks such as object detection and localization.
Essential applications are traffic sign classification or detection \cite{tabernik2019deep, DBLP:conf/date/ChengHBH20}, lane detection \cite{li2016deep}, vehicle or pedestrian detection \cite{chen2021deep}, driver monitoring and driver-vehicle interaction\cite{diederichs2022artificial}.
All these functions are safety-critical, because incorrect outputs may create dangerous situations, accidents and even loss of life.
Therefore, testing them for security, reliability, and robustness has the utmost priority before deploying the functions on autonomous vehicles into the real world. 

DNN unfortunately can easily be manipulated due to their dependency on the training data. For example, consider a traffic sign classification model trained on a large dataset such as GTSRB~\cite{stallkamp2012man}.
An attacker having access to the data during training may intentionally poison it by modifying a small percentage of the data.
This can be done by adding \emph{trojan patterns} to the input belonging to different classes.
The \emph{trojan patterns} may be in the form of objects, image transformations, invisible watermarks and many more. The model trained with such poisoned data may have learned false features called \emph{backdoors} which have no direct relation to the classification output.
Such models still perform well on benign inputs; however, they may fail in the presence of trojan patterns (which only the attackers know).

Research has shown that backdoors may exist even on models trained with benign data \cite{liu2019abs}. 
This is because certain features may have strong correlation to an output class making the model biased towards such features.
For example, traffic signs such as \emph{pedestrian crossing} may usually have urban background whereas \emph{wild animal crossing} may usually have country/rural backgrounds.
In such cases, the DNN may have learned the background instead of the traffic sign itself leading to bias and in turn misclassification.
Therefore, it is vital to defend against both intentional backdoors (present due to an attacker's poisoning of training data) and unintentional backdoors (present due to a strong correlation to certain features for a few classes) to ensure the proper functionality of machine learning models.

\emph{Coverage testing} is one of the typical software testing approaches where the goal is to achieve complete code coverage by checking the correctness for the entire input space.
Using such techniques to test DNN is not straightforward, due to the massive number of parameters and the black-box nature of DNN.
However, there has been a vast amount of research in adapting those coverage techniques to work with DNN.
One such approach is \emph{NN-dependability}~\cite{cheng2018towards}, which proposes several metrics to measure quality of the DNN in terms of robustness, interpretability, completeness, and correctness.
However, the metrics cannot test for backdoors.
Other software engineering techniques such as \emph{Modified Condition/Decision MC/DC}~\cite{sun2019structural} and \emph{scenario based testing}~\cite{cheng2018quantitative} also do not focus on security aspects such as backdoor testing.
Our approach is different from these as we target specifically at overcoming backdoors and biases in the DNN.

\begin{figure}[t]
    \includegraphics[width=12.0cm, trim={0.0cm 0.0cm 0.0cm 0.0cm},clip]{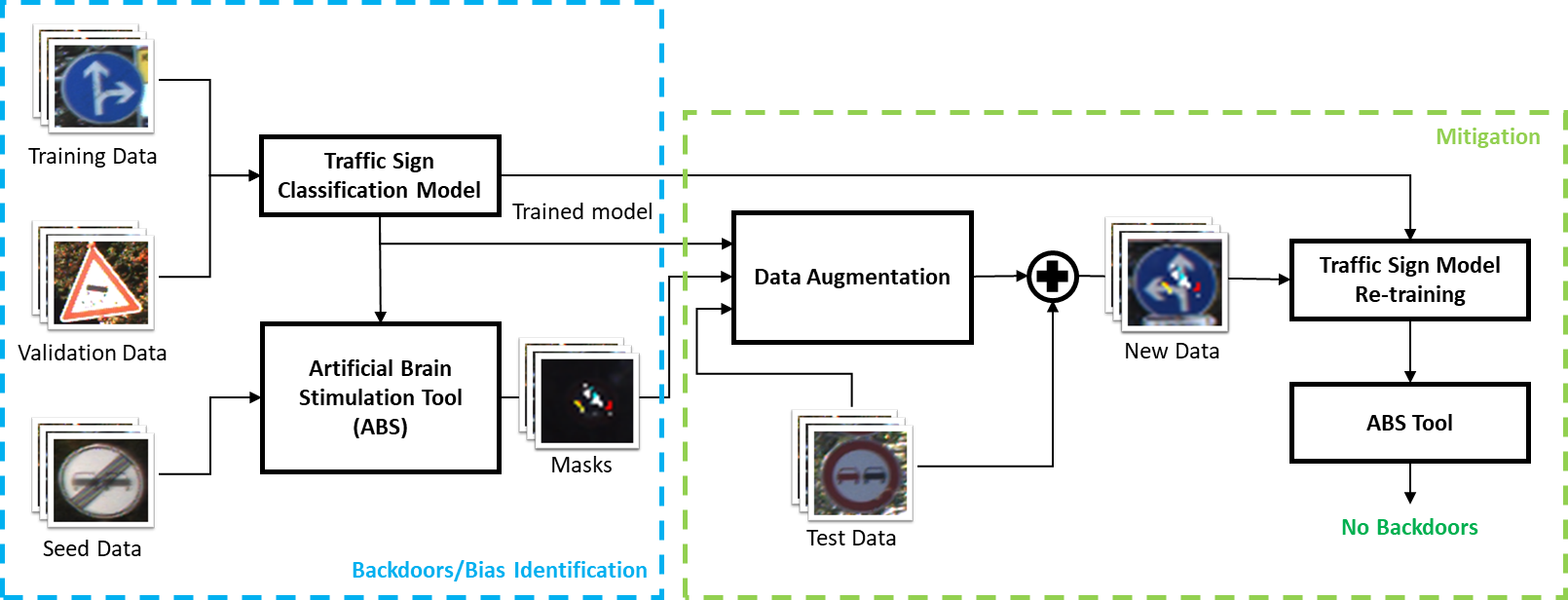}
    \caption{Framework of the backdoor or bias mitigation approach} 
    \vspace{-0.50cm}
    \label{Fig: Framework}
\end{figure}

Several attacking techniques developed in recent years \cite{nguyen2020wanet, doan2021lira, nguyen2020input, liu2020reflection} are excellent at fooling even the state-of-the-art defense methods such as \emph{STRIP}~\cite{gao2019strip}, \emph{Fine-pruning}~\cite{liu2018fine}, and \emph{Neural Cleanse}~\cite{wang2019neural}.
It is essential to defend from such attacks, especially for safety-critical applications.
A defense mechanism includes two phases.
The first phase is to detect the backdoors and the second is to mitigate them.
Detection techniques such as \cite{wang2019neural, liu2018fine} can identify common kinds of poisoning such as masking with patches, noise and watermarks. 
However, they are \emph{white-box}, meaning that they need information about the type or position of trojan patterns.
The detection technique needs to be able to treat the data as a black box, because we usually do not have any information on how the data is poisoned \cite{liu2019abs, dong2021black}.

The second step is \emph{mitigation} where we utilize the outputs from the detection techniques and modify the DNN to defend against attacks.
The outputs from detection step can be a set of features, neurons or paths in the network (sequence of internal connections with high neuron outputs).
Mitigation techniques focus on modifying the inner parameters such as \emph{neuron repair}~\cite{yang2022neural} where unsafe regions are detected and repaired post-hoc, \emph{anti-backdoor learning}~\cite{li2021anti} where effectiveness of the poisoned data is limited by controlling the learning speed during training.
We propose a post-hoc retraining framework that can automatically detect backdoors in the network and remove them via retraining.
Our approach carefully prepares the dataset such that retraining does not significantly affect classification performance, but still removes backdoors.

Figure~\ref{Fig: Framework} depicts our approach in a nutshell.
We utilize a black-box backdoor identification technique called Artificial Brain Stimulation (ABS) by Liu \etal~\cite{liu2019abs}. 
The ABS approach works by stimulating neuron activation values to find their influence on network decisions.
A neuron is highly influential or \emph{poisoned} if a change in the activation value of a neuron shifts the DNN classification output to a different class.
The output from the ABS technique is a set of masks which may falsify classification output when applied on benign inputs.
We utilize these masks to remove backdoors from the DNN model.
The overview of our mitigation approach is on the right side of Figure~\ref{Fig: Framework}.
Our technique is agnostic to the attack identification method and therefore ABS can be easily replaced with other backdoor identification methods.
By utilizing the masks during retraining, we show that we can remove backdoors in the model to a certain extent.

Our approach shares some ideas with \emph{Neural Cleanse}~\cite{wang2019neural} where they employ backdoor mitigation via \emph{unlearning}, meaning that they retrain the DNN model using a small percentage of training data combined with the masked data.
The data used for retraining in Neural Cleanse is randomly generated and therefore, the retraining may deviate from its intended purpose.
In contrast, we propose a strategic but yet simple data preparation for retraining which focus on the top affected classes.
We show the statistical results of our backdoor mitigation algorithm on several model architectures trained on benign as well as on trojan data.

\section{Preliminaries}
\label{Sec: Preliminaries}
This section briefly introduces DNN and the types of networks considered in the paper. Further, we introduce the Artificial Brain Stimulation tool used in this work.

\subsection{Deep Neural Networks}
This work focuses on the classification problem and thus uses a simple architecture with convolutional layers.
We represent a \emph{Deep Neural Network} as a tuple, $\mathcal{N}=(\mathbb{S}, \mathbb{T}, \phi)$, where $\mathbb{S} = \{\mathbb{S}_k | k \in \{1, \ldots, K\}\}$ is a set of layers with $K$ being the total number of layers, $\mathbb{T} \subseteq \mathbb{S} \times \mathbb{S}$ is a set of connections between the layers and $\phi =  \{\phi_k | k \in \{2, \ldots, K\}\}$ is a set of functions, one for each non-input layer.
A typical DNN has an input layer $\mathbb{S}_1$, an output layer $\mathbb{S}_K$ and several \emph{hidden layers} between the input and the output.
Each layer $k$ consists of $S_k$ number of neurons/nodes. 
Let us define the $l$-th neuron of layer $k$ as $n_{k,l} \in \mathbb{S}_k$.
Each neuron $n_{k,l}$ for $2 \leq k \leq K-1$ and $1 \leq l \leq S_K$ is associated with a value before activation $u_{k,l}$ and a value after activation $v_{k,l}$.
The activation is a function that modifies the input based on a formula.
We use the Rectified Linear Unit (ReLU) activation function in this work.

In a classification model, the output dimension or number of neurons in the output layer $S_K$ is equal to the number of labels $\mathcal{L} = \{1,\ldots,S_K\}$, which means the classification output defined as $\mathit{label} = \text{argmax}_{1 \leq l \leq S_K} u_{K,l}$ is the index of the neuron in the output layer with the largest value.
We define input data as $X = \{x_1, \ldots, x_T \}$
where each $x_i$ is an image that is passed to the DNN.
The classification output for an input $x$ is denoted as $\mathcal{N}[x]$.
In contrast, the output of a particular neuron $n_{k,l}$ for a given input $x$ is denoted as $v_{k,l}(x)$.

\subsection{Artificial Brain Stimulation Analysis}
\label{Sec: ABS}
Artificial Brain Stimulation Analysis aims to identify backdoors in a trojan or benign model.
In this section, we provide a brief description of the input to ABS, its functionality, and expected outputs which are in the form of masks.
The input to the model is a trained DNN
$\mathcal{N}$. We also require seed data $X_\mathit{seed} = \{x_1,\ldots,\,x_T\}$ where $T \geq S_K$ and $\{\forall t\!\in\!\mathcal{L} \ \exists x \! \in \! X_\mathit{seed} \ s.t \ \mathcal{N}[x] \! = \! t \}$ meaning a set of benign images with at least one associated to each class.
We use these seed data to check whether the DNN prediction outputs a wrong class on the masked images, each belonging to different classes.
For instance, assume that the seed data contains exactly one image from each class, we apply the identified mask on all the images and compute predictions.
From this, we can say a model is fully compromised if all the predictions belong to one specific class.

The ABS analysis has three steps.
The first step is to perform \emph{stimulation analysis} where we replace the activation value $v_{k,l}$ of the neuron under analysis $n_{k,l}$ with the stimulation value $z_{k,l}$.
We do such analysis for each neuron $n_{k,l} \in \mathbb{S}_k$ in all hidden layer $2 \leq k \leq K-1$.
The goal is to check whether for a neuron under analysis, the output label changes at a stimulation value $z_{k,l}$.
As a result, we obtain the \emph{neuron stimulation function} (NSF) which provides the output class $i \in \mathcal{L}$ for different stimulation values $z_{k,l}$.
Note that, during stimulation analysis of the $l^\mathit{th}$ neuron in layer $k$, the values of the rest of the neurons in that layer $k$ do not change.
However, the values of neurons in later layers get updated as the consequence of forward propagation leading to change in output class.
We refer readers to the original paper \cite{liu2019abs} for more details on the stimulation procedure.

\begin{algorithm}[!htb]
    \caption{Backdoor mitigation via retraining}
    \begin{algorithmic}[1]
        \renewcommand{\algorithmicrequire}{\textbf{Input:}}
        \renewcommand{\algorithmicensure}{\textbf{Output:}}
        \REQUIRE $\mathcal{N}$: Trained DNN,\\
        $M_\mathit{masks}$: trojan masks from ABS analysis on $\mathcal{N}$,\\
        $X_\mathit{seed}$: seed data for ABS analysis on retrained model,\\
        $X_\mathit{test} = \{x_1, \cdots, x_T \}$: benign test data,\\
        $y_\mathit{test} = \{y_1, \cdots, y_T \}$: true labels for data augmentation,\\
        $X_\mathit{valid}$: benign validation data to track the drop in accuracy,\\
        $\mathit{top}_p$: parameter to control the number of classes considered for new data generation,\\
        $\delta$: accuracy drop threshold.
        \ENSURE $\hat{\mathcal{N}}$: Retrained DNN without backdoors or bias.
        % \\ \textit{Initialisation} :
        \STATE Initialize $\hat{\mathcal{N}}$ with learned weights from the network $\mathcal{N}$. \label{Line: Initialize DNN}
        \WHILE {(accuracy of $\hat{\mathcal{N}}$ - accuracy of $\mathcal{N}$ on $X_\mathit{valid}$) $\leq \delta$} \label{Line: While Start}
        \STATE Initialize $X_\mathit{new}$ and $y_\mathit{new}$ as re-training data and true labels respectively.
        \FOR {Mask in $M_\mathit{masks}$} \label{Line: Step 1}
        \STATE Define $X_\mathit{test}'$ as images after applying masks on test data.
        \STATE Let $y_\mathit{test}'$ be the according predictions.
            \FOR {Img, label in $X_\mathit{test}, y_\mathit{test}$}
                \STATE Apply $\mathit{mask}$ on $\mathit{img}$.
                \STATE Predict $\hat{\mathcal{N}}[\textrm{masked image}]$.
                \STATE Add the masked image and prediction to $X_\mathit{test}'$ and $y_{test}'$.
            \ENDFOR 
            \STATE Compute False Positives using $y_\mathit{test}'$ and $y_\mathit{test}$. \label{Line: false positives}
            \STATE Select $\mathit{top}_p$ number of classes with the highest false positives.
            \STATE Update $X_\mathit{new}$ with all false positive images belonging to $\mathit{top}_p$ classes.
            \STATE Update $y_\mathit{new}$ with respective true labels. \label{Line: update new labels}
        \ENDFOR \label{Line: Step 1 End}
        \STATE $X_\mathit{new}$.extend($X_\mathit{test}$)
        \STATE $y_\mathit{new}$.extend($y_\mathit{test}$)  
        \STATE Shuffle and Split $X_\mathit{new}$ and $y_\mathit{new}$ as training and validation dataset. \label{Line: Step 2}
        \STATE Retrain $\hat{\mathcal{N}}$ with new training and validation dataset. \label{Line: Retrain}
        \STATE Analyze $\hat{\mathcal{N}}$ using ABS tool to identify backdoors $\hat{X}_\mathit{masks}$. \label{Line: Step 3}
        \IF {$\hat{X}_\mathit{masks} = \emptyset$}
            \RETURN DNN $\hat{\mathcal{N}}$. \label{Line: Return}
        \ENDIF
        \ENDWHILE \label{Line: While Stop}
    \end{algorithmic}
    \label{Alg: Mitigation}
\end{algorithm}

The next step is to find a set of \emph{compromised neurons} using the NSFs.
A neuron $n_{k,l}$ is said to be \emph{compromised} if, for the stimulation value falling in a particular range, the outputs of all NSFs generated from the seed data respectively are same.
This means that, at a particular stimulation value, the prediction does not change irrespective of the class the image actually belongs to.
Let us define $C$ as the total number of such candidates.

The last step is to obtain masks for each compromised neuron via \emph{reverse engineering}.
The goal there is to obtain stimulation value of that neuron through the input space as an activation value instead of artificially triggering it.
Therefore, we obtain masks denoted as $M = \{m_1, \cdots, m_C\}$ for each compromised neuron candidate.
Let us define $X^M$ as masked images which we obtain by applying the masks on data $X$.
Further, we define the \emph{Attack Success Rate} (ASR) as the percentage of misclassification on the masked images $X^M$.
Using these, we set a threshold parameter denoted as \emph{REASR bound} which is based upon ASR on masked images $X_\mathit{seed}^M$ and therefore ranges between $0$ to $1$.
The REASR bound will filter the masks that affect very few classes.
Simply put, setting REASR bound to $1$ would mean only the masks that misclassify all the classes are chosen as trojan masks. 
After filtering, we obtain the final trojan masks denoted as $M_\mathit{masks} = \{m_1, \cdots, m_M\}$.

\section{Methodology}
\label{Sec: Methodology}
In this work, our goal is to eradicate backdoors in the DNN model by retraining.
Algorithm~\ref{Alg: Mitigation} illustrates our approach.
We require a trained DNN model $\mathcal{N}$, masks $M_\mathit{masks}$ from the ABS analysis and benign test data $X_\mathit{test}$, $y_\mathit{test}$.
Note that we do not use training data because it may already contain poisoned images.
The expected output from this algorithm is a benign DNN model $\hat{\mathcal{N}}$ with no backdoors.

This method has three main steps as also depicted in the green box highlighted in Figure~\ref{Fig: Framework}.
The first step is the data augmentation in lines~\ref{Line: Step 1} $-$ \ref{Line: Step 1 End}. For each mask, we apply the mask on all the test data and obtain their predictions on the DNN $\hat{\mathcal{N}}$.
Next, we compute the \emph{confusion matrix} to obtain the false positives for each class as in line~\ref{Line: false positives}.
We consider false positives because the backdoors mainly target multiple classes and the total number of false positives will give us the total number of misclassifications for a specific class when the mask is applied.
Our strategy is to consider $\mathit{top}_p$ classes with the highest number of false positives for a given dataset so that the retraining will focus more on those highly affected classes.
We add the images from this $\mathit{top}_p$ classes that were wrongly classified to our new dataset $X_\mathit{new}$ as in line~\ref{Line: false positives} $-$ \ref{Line: update new labels}.
Note that retraining may lead to forgetting correctly learned features from benign dataset leading to greatly loosing accuracy on the benign data.
To overcome this, we combine $X_\mathit{new}$ with benign $X_\mathit{test}$ data so that retaining would not overfit towards the new data $X_\mathit{new}$.

In the next two steps, we utilize $X_\mathit{new}$ to retrain DNN $\hat{\mathcal{N}}$ in line~\ref{Line: Retrain} and then analyze the model for backdoors using ABS tool in line~\ref{Line: Step 3}.
If backdoors are found, we repeat the steps in lines~\ref{Line: While Start} $-$ \ref{Line: While Stop}.
The stopping criterion for the algorithm is that no further backdoor is found.
In this case, we return the DNN $\hat{\mathcal{N}}$ as in line~\ref{Line: Return}.
On the other hand, we set a threshold $\delta$ as another stopping criteria to check the drop in accuracy of the new DNN $\hat{\mathcal{N}}$ on $X_\mathit{valid}$ and stop retraining when the accuracy drop goes below it.
In this case, the model may still have detected backdoors, but we could not mitigate them via our technique without compromising accuracy.

\section{Experiments}
In this section, we show the results of performing backdoor mitigation.
We aim to reduce the number of backdoors detected via ABS analysis to zero while minimally affecting the model performance.
In order to do so, the trojan model has to unlearn the \emph{poisoned patterns} to avoid safety and security risks during deployment.
We show that our idea of targeting the $\mathit{top}_p$ classes for retraining the model helps to remove biases without compromising performance.
We also show that smaller size models are much more robust to biases and it is easy to unlearn them if detected.
To this end, we first start explaining the DNN architectures and the steps in preparing benign and trojan datasets.
Next, we show the results from performing ABS analysis on the DNN models.
Finally, we show the experimental results from the mitigation algorithm presented in Section~\ref{Sec: Methodology}.

\subsection{Experiment Setup}
The focus of this section is to briefly describe the experimental setup to evaluate our approach. 
Precisely, the results in this section are from the backdoor/bias identification phase in the framework~\ref{Fig: Framework}.
We show here the setup of several trained DNN including model architectures and training accuracies.
Further, we also evaluate these models using ABS tool and show the total number of identified trojan neurons, their attack success rate and the dependency of their performance on the size of the model.

\begin{figure}[htbp]
    \vspace{-0.25cm}
    \includegraphics[width=\textwidth]{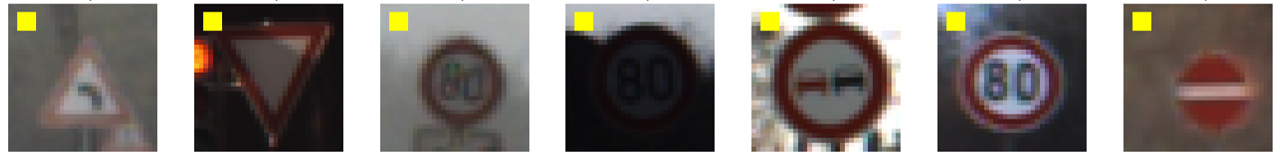}
    \caption{Sample of trojaned images} 
    \label{Fig: Trojaned Data}
    \vspace{-0.25cm}
\end{figure}

In this work, we utilize the GTSRB dataset \cite{stallkamp2012man} for traffic sign classification.
We split the dataset into four parts:
$(X_\mathit{train}, y_\mathit{train})$\footnote{For simplicity, the label $y$ is emitted from the text in the upcoming descriptions; however it exists unless specifically stated otherwise} with size $35228$, $X_\mathit{valid}$ with size $4410$, $X_\mathit{test}$ with size $12630$, and $X_\mathit{seed}$ with size $43$.
Additionally, we develop a trojan dataset $X_\mathit{train}^\mathit{troj}$, $X_\mathit{valid}^\mathit{troj}$ by adding yellow patches to $20\%$ of the images in both $X_\mathit{train}$ and $X_\mathit{valid}$ and modify all their labels to target to one unique class.
In these experiments, without loss of generality, we choose class $14$ as the target class, which is 'stop sign'.
Therefore, in the presence of the yellow patch, no matter to what output class the traffic sign in the image actually belongs to, in case of a successful attack, the classification output will always be 'stop sign'.
A sample of trojan images is depicted in Figure~\ref{Fig: Trojaned Data}.

\begin{table}[htbp]
    \vspace{-0.25cm}
    \setlength{\tabcolsep}{4pt} 
    \centering
    \caption{Model architecture and training information}
    \begin{tabular}{lccc}
        \toprule
         & $\mathcal{N}_\mathit{SN}$ & $\mathcal{N}_\mathit{MN}$ & $\mathcal{N}_\mathit{LN}$ \\
        \midrule
        Model architecture & $4$ Conv $+$ 1 Dense & $5$ Conv $+$ 1 Dense & $5$ Conv $+$ 1 Dense \\
        Features in each layer & $[8, 16, 32, 16]$ & $[16, 32, 64, 32, 16]$ & $[32, 64, 128, 64, 32]$ \\
        Trainable parameters & $30203$ & $130091$ & $516139$ \\
        \bottomrule
    \end{tabular}
    \vspace{-0.25cm}
    \label{Table: Model Arcitcture}
\end{table}

We train three DNN models using benign dataset $X_\mathit{train}$, $X_\mathit{valid}$ and call them \emph{small size} $\mathcal{N}_\mathit{SN}$, \emph{moderate size} $\mathcal{N}_\mathit{MN}$, and \emph{large size network} $\mathcal{N}_\mathit{LN}$.
These three networks have similar architectures with variable layers and features as depicted in Table~\ref{Table: Model Arcitcture}.
Similarly, we train three trojan models $\mathcal{N}_\mathit{SN}^\mathit{troj}$, $\mathcal{N}_\mathit{MN}^\mathit{troj}$, $\mathcal{N}_\mathit{LN}^\mathit{troj}$ using trojaned dataset $X_\mathit{train}^\mathit{troj}$, $X_\mathit{valid}^\mathit{troj}$.
Table~\ref{Table: Model Accuracies} depicts the classification accuracies of all these models.

\begin{table}[htbp]
    \setlength{\tabcolsep}{4pt} 
    \centering
    \caption{Accuracies of the trained model}
    \begin{tabular}{lllllll}
        \toprule
        \multirow{2}{*}[-0.25em]{Dataset} & \multicolumn{3}{c}{Benign Models} & \multicolumn{3}{c}{Trojan Models}\\
        \cmidrule(lr){2-4} \cmidrule(lr){5-7} \\
        \addlinespace[-8pt]
        {} & $\mathcal{N}_\mathit{SN}$ & $\mathcal{N}_\mathit{MN}$ & $\mathcal{N}_\mathit{LN}$ & $\mathcal{N}_\mathit{SN}^\mathit{troj}$ & $\mathcal{N}_\mathit{MN}^\mathit{troj}$ & $\mathcal{N}_\mathit{LN}^\mathit{troj}$ \\
        \midrule
        Training data $X_\mathit{train}$ & $98.80\%$ & $99.45\%$ & $99.24\%$ & $99.30\%$ & $99.55\%$ & $99.52\%$ \\
        Validation data $X_\mathit{valid}$ & $91.75\%$ & $94.29\%$ & $95.35\%$ & $92.61\%$ & $93.51\%$ & $96.44\%$ \\
        Test data $X_\mathit{test}$ & $87.99\%$ & $90.10\%$ & $91.53\%$ & $87.93\%$ & $91.54\%$ & $91.94\%$ \\
        \bottomrule
    \end{tabular}
\vspace{-0.50cm}
    \label{Table: Model Accuracies}
\end{table}

Next we run the ABS analysis on these models and generate masks $M_\mathit{masks}$.
We set the parameters of the ABS tool similar to the authors \cite{liu2019abs} except \emph{REASR bound} which is set to $0.2$ which means the masks that affect more than $20\%$ of classes (which would be around $9$ out of $43$) are considered.
The reason to set this to $0.2$ is to control the number of trojan masks.
It is worthwhile to mention that setting the REASR bound to higher values will not output any trojan masks and setting them to lower values will output many masks that are however less effective.

Finally, we apply these masks on the test data $X_\mathit{test}$ to obtain a new set of masked images $X_\mathit{test}^M$ and afterwards compute model predictions on them.
Table~\ref{Table: ABS Analysis} shows the number of trojan neurons, and ASR on masked images.
Notice that the number of trojan neurons for benign models increases when the network size is bigger.
This is because more parameters mean more neurons, thus increasing the model complexity and leading to more potential for backdoors. 
The attack success rate of trojan models on $X_\mathit{test}^M$ is large because the ABS tool successfully found the imputed trojan pattern.
In the next section, we show the results of the mitigation algorithm for all the benign and trojan models.

\begin{table}[htbp]
    \vspace{-0.25cm}
    \setlength{\tabcolsep}{6pt} 
    \centering
    \caption{Results from ABS analysis which includes number of trojan neurons and attack success rates on respective $X_\mathit{seed}^M$ data}
    \begin{tabular}{lllllll}
        \toprule
        \multirow{2}{*}[-0.25em]{Property} & \multicolumn{3}{c}{Benign Models} & \multicolumn{3}{c}{Trojan Models}\\
        \cmidrule(lr){2-4} \cmidrule(lr){5-7} \\
        \addlinespace[-8pt]
        {} & $\mathcal{N}_\mathit{SN}$ & $\mathcal{N}_\mathit{MN}$ & $\mathcal{N}_\mathit{LN}$ & $\mathcal{N}_\mathit{SN}^\mathit{troj}$ & $\mathcal{N}_\mathit{MN}^\mathit{troj}$ & $\mathcal{N}_\mathit{LN}^\mathit{troj}$ \\
        \midrule
        $\#$ of Trojan Neurons & $1$ & $3$ & $3$ & $4$ & $3$ & $2$\\
        Attack Success Rate & $67.43\%$ & $76.46\%$ & $70.86\%$ & $97.69\%$ & $93.00\%$ & $80.72\%$\\
        \bottomrule
    \end{tabular}
    \label{Table: ABS Analysis}
    \vspace{-0.75cm}
\end{table}

\subsection{Mitigation Results}

\newcommand{\width}{3.0cm}
\begin{figure}[!htbp]
    \centering
    \subfigure{
        \begin{tikzpicture}
            \begin{axis}[
                xlabel={Predictions},ylabel={Actual Label},
                xtick distance=5,ytick distance=5,
                xtick pos=bottom, ytick pos=left,
                label style={font=\scriptsize},
                tick label style={font=\scriptsize, /pgf/number format/fixed},
                axis on top, % ----
                width=\width,
                height=\width,
                scale only axis=true,
                enlargelimits=false,
                tick align=outside,
                ]
            \addplot graphics[xmin=0, xmax=43, ymin=0.0, ymax=43, includegraphics={width=\width, trim={1.55cm 1.45cm 0.25cm 0.25cm},clip}] {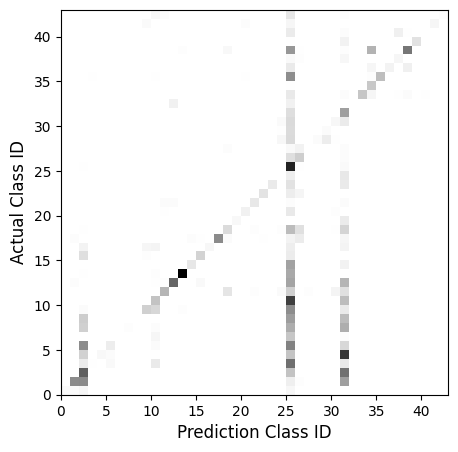};
            \end{axis}
        \end{tikzpicture}
    }
    \subfigure{
        \begin{tikzpicture}
            \begin{axis}[
                xlabel={Predictions},
                xtick distance=5,ytick distance=5,
                xtick pos=bottom, ytick pos=left,
                label style={font=\scriptsize},
                tick label style={font=\scriptsize, /pgf/number format/fixed},
                axis on top, % ----
                width=\width,
                height=\width,
                scale only axis=true,
                enlargelimits=false,
                tick align=outside,
                ]
            \addplot graphics[xmin=0, xmax=43, ymin=0, ymax=43, includegraphics={width=\width, trim={1.55cm 1.45cm 0.25cm 0.25cm},clip}] {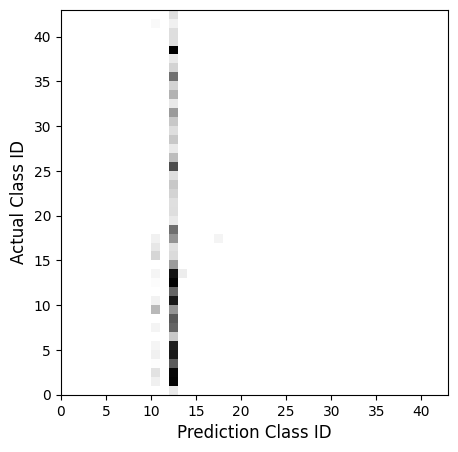};
            \end{axis}
        \end{tikzpicture}
    }
    \subfigure{
        \begin{tikzpicture}
            \begin{axis}[
                xlabel={Predictions},
                xtick distance=5,ytick distance=5,
                xtick pos=bottom, ytick pos=left,
                label style={font=\scriptsize},
                tick label style={font=\scriptsize, /pgf/number format/fixed},
                axis on top, % ----
                width=\width,
                height=\width,
                scale only axis=true,
                enlargelimits=false,
                tick align=outside,
                ]
            \addplot graphics[xmin=0, xmax=43, ymin=0.0, ymax=43, includegraphics={width=\width, trim={1.55cm 1.45cm 0.25cm 0.25cm},clip}] {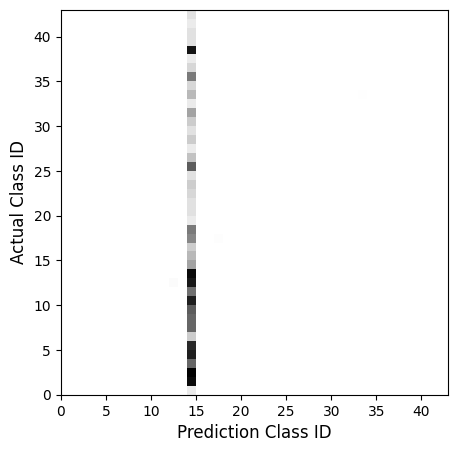};
            \end{axis}
        \end{tikzpicture}
    }
    \subfigure{
        \hspace{3.85cm}
        \begin{tikzpicture}
            \begin{axis}[
                xlabel={Predictions},ylabel={Actual Label},
                xtick distance=5,ytick distance=5,
                xtick pos=bottom, ytick pos=left,
                label style={font=\scriptsize},
                tick label style={font=\scriptsize, /pgf/number format/fixed},
                axis on top, % ----
                width=\width,
                height=\width,
                scale only axis=true,
                enlargelimits=false,
                tick align=outside,
                ]
            \addplot graphics[xmin=0, xmax=43, ymin=0, ymax=43, includegraphics={width=\width, trim={1.55cm 1.45cm 0.25cm 0.25cm},clip}] {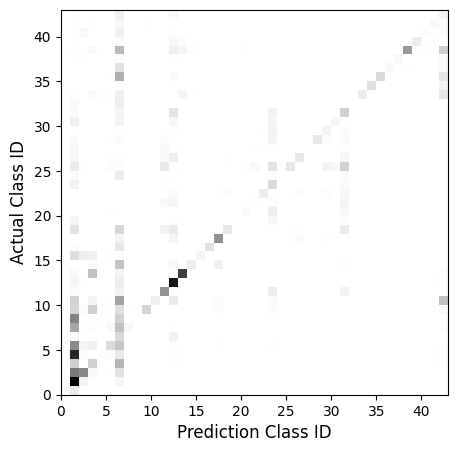};
            \end{axis}
        \end{tikzpicture}
    }
    \subfigure{
        \begin{tikzpicture}
            \begin{axis}[
                xlabel={Predictions},
                xtick distance=5,ytick distance=5,
                xtick pos=bottom, ytick pos=left,
                label style={font=\scriptsize},
                tick label style={font=\scriptsize, /pgf/number format/fixed},
                axis on top, % ----
                width=\width,
                height=\width,
                scale only axis=true,
                enlargelimits=false,
                tick align=outside,
                ]
            \addplot graphics[xmin=0, xmax=43, ymin=0, ymax=43, includegraphics={width=\width, trim={1.55cm 1.45cm 0.25cm 0.25cm},clip}] {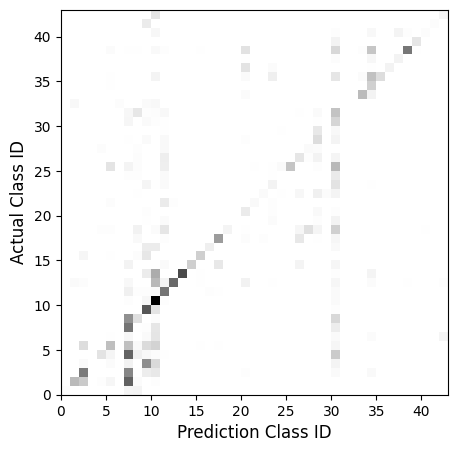};
            \end{axis}
        \end{tikzpicture}
    }
    \caption{Confusion Matrix from predictions of model $\mathcal{N}_\mathit{SN}$ on data $X_\mathit{test}^{M_1}$ (image in first column from left) and predictions of model $\mathcal{N}_\mathit{SN}^\mathit{troj}$ on data $X_\mathit{test}^{M_2}$ (images in second and third columns).}
    \vspace{-0.40cm}
    \label{Fig: Confusion Matrix}
\end{figure}

Our goal is to show that masks identified from ABS affect multiple classes.
For this, we utilize confusion matrices depicted in Figure~\ref{Fig: Confusion Matrix}, which we obtain using the actual labels $y_\mathit{test}$ and predictions from model $\mathcal{N}_\mathit{SN}$ on data $X_\mathit{test}^{M_1}$ where $M_1 = \{m_1\}$ (data by applying one mask from ABS analysis) and from model $\mathcal{N}_\mathit{SN}^\mathit{troj}$ on data $X_\mathit{test}^{M_2}$ where $M_2 = \{m_1, m_2, m_3, m_4\}$ (data by applying three masks from ABS analysis).
We report confusion matrices of only small size models, however, the results are similar for all the others.
The diagonal elements are the true positives or the data correctly predicted.
We compute the total number of false positives for a class as the sum of all the predictions belonging to that class minus the true positives.
The multiple columns with high color intensities in Figure~\ref{Fig: Confusion Matrix} show that benign and trojan models may have a backdoor affecting more than one class.
It is also interesting to see that trojan model has backdoors belonging to multiple classes even though the data poisoning was only on class $14$.

\begin{figure}[htbp]
    \vspace{-0.50cm}
    \centering
    \subfigure{
        \includegraphics[width=5.5cm, trim={0.0cm 0.0cm 0.0cm 0.0cm},clip]{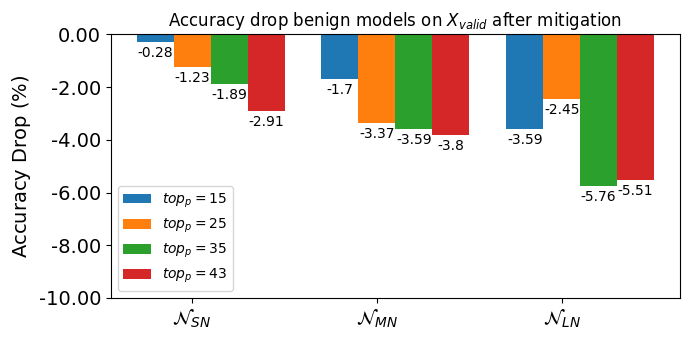}}
    \subfigure{
        \includegraphics[width=5.5cm, trim={0.0cm 0.0cm 0.0cm 0.0cm},clip]{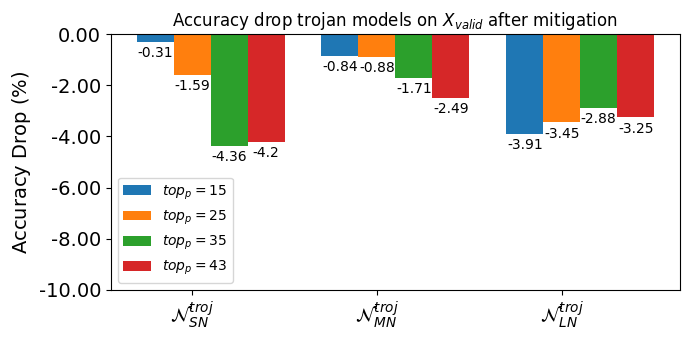}}
    \caption{Drop in classification accuracy after retraining at different $\mathit{top}_p$ values}
    \vspace{-0.45cm}
    \label{Fig: Accuracy drop}
\end{figure}

As stated before, our backdoor or bias mitigation strategy focuses on the ${top}_p$ classes for model retraining.
Therefore, we run four experiments for each trained model by setting $\mathit{top}_p$ to $15$, $25$, $35$ and $43$, respectively.
Figure~\ref{Fig: Accuracy drop} depicts the drop in accuracy after running the algorithm.
We utilize benign validation data $X_\mathit{valid}$ to check the drop in accuracy for both benign and trojan models.
As we can see, for both types of models, the drop in accuracy strongly depends on $\mathit{top}_p$ value.
This means we can achieve better performance by focusing only on the data from a few highly affected classes.

Table~\ref{Table: Mitigation Iterations} shows the change in the number of trojan neurons and attack success rate after retraining once.
Observe the drop in the respective ASRs when we restrict retraining to smaller $\mathit{top}_p$. 
The advantage of retraining with smaller $\mathit{top}_p$ is that we can mitigate backdoors better by considering only top-affected classes without losing the classification performance of the DNN. 
To show the effectiveness of our method, we train another trojan model $\mathcal{N}_\mathit{NCN}^\mathit{troj}$ with the same architecture and trojaning technique as in Neural Cleanse.
Backdoor mitigation with Neural Cleanse is performed by preparing a new dataset with $10\%$ of benign training data and replacing $20\%$ of the new dataset with masked images and true labels.
The network is then trained for only $1$ epoch.
We show the comparison results in Table~\ref{Table: Comparison results} where we can see that we are able to achieve much lower attack success rate without affecting the classification accuracy.

\begin{table}[htbp]
    \vspace{-0.25cm}
    \setlength{\tabcolsep}{5pt} 
    \centering
    \caption{Number of detected trojan neurons and their attack success rate after retraining once}
    \begin{tabular}{lcccccccccc}
        \toprule
        \multirow{2}{*}[-0.25em]{Model} & \multicolumn{5}{c}{$\#$ of trojan neurons} & \multicolumn{5}{c}{Attack success rate}\\
        \multirow{2}{*}[-0.25em]{} & \multicolumn{5}{c}{at different $\mathit{top}_p$ values} & \multicolumn{5}{c}{at different $\mathit{top}_p$ values}\\
        \cmidrule(lr){2-6} \cmidrule(lr){7-11} \\
        \addlinespace[-8pt]
        {} & Before & $43$ & $35$ & $25$ & $15$ & Before & $43$ & $35$ & $25$ & $15$ \\
        \midrule
        $\mathcal{N}_\mathit{SN}$ & $1$ & $1$ & $1$ & $0$ & $0$ & $67.43\%$ & $40.60\%$ & $35.00\%$ & $0.0\%$ & $0.0\%$ \\
        $\mathcal{N}_\mathit{MN}$ & $3$ & $1$ & $2$ & $1$ & $0$ & $76.46\%$ & $90.01\%$ & $84.67\%$ & $64.82\%$ & $0.0\%$ \\
        $\mathcal{N}_\mathit{LN}$ & $3$ & $2$ & $1$ & $0$ & $0$ & $70.86\%$ & $68.02\%$ & $64.02\%$ & $0.0\%$ & $0.0\%$ \\
        \midrule
        $\mathcal{N}_\mathit{SN}^\mathit{troj}$ & $4$ & $3$ & $3$ & $1$ & $0$ & $97.69\%$ & $91.60\%$ & $80.27\%$ & $62.73\%$ & $0.00\%$ \\
        $\mathcal{N}_\mathit{MN}^\mathit{troj}$ & $3$ & $2$ & $2$ & $1$ & $1$ & $93.00\%$ & $92.74\%$ & $64.20\%$ & $71.42\%$ & $38.30\%$ \\
        $\mathcal{N}_\mathit{LN}^\mathit{troj}$ & $2$ & $1$ & $1$ & $0$ & $0$ & $80.72\%$ & $90.79\%$ & $43.05\%$ & $0.0\%$ & $0.0\%$ \\
        \bottomrule
    \end{tabular}
    \vspace{-0.40cm}
    \label{Table: Mitigation Iterations}
\end{table}

We show the number of trojan neurons after retraining multiple times in Table~\ref{Table: Multiple Mitigation} with the maximum drop in accuracy $\delta$ set to $8\%$.
It is worth mentioning that the drop in accuracy after three iterations for smaller networks is at most five percent, but we set $\delta$ to $8\%$ so that all the networks can be retrained at least twice (see Figure~\ref{Fig: Accuracy drop}).
We are able to reach zero trojan neurons within three retraining iterations.
Notice that setting higher $\mathit{top}_p$ values may sometime increase the number of trojan neurons in the network.
On the other hand, lower $\mathit{top}_p$ values can remove all trojan neurons in fewer iterations making our mitigation technique very effective.

\begin{table}[htbp]
    \vspace{-0.25cm}
    \setlength{\tabcolsep}{4.pt} 
    \centering
    \caption{Mitigation comparison with Neural Cleanse on model $\mathcal{N}_\mathit{NCN}^{troj}$}
    \begin{tabular}{lcc}
        \toprule
        Mitigation method & Classification Accuracy & Attack Success Rate \\
        \midrule
        Before Mitigation & $97.27\%$ & $96.45\%$ \\
        \midrule
        Neural Cleanse & $94.25\%$ & $19.18\%$ \\
        Our Approach & $\bm{95.77\%}$ & $\bm{5.38\%}$ \\
        \bottomrule
    \end{tabular}
    \vspace{-0.40cm}
    \label{Table: Comparison results}
\end{table}

\begin{table}[!htbp]
    \setlength{\tabcolsep}{5pt} 
    \centering
    \caption{Number of trojan neurons at different $\mathit{top}_p$ values and at different mitigation iterations}
    \begin{tabular}{lcccccccccccc}
        \toprule
        Model & \multicolumn{3}{c}{$\mathit{top}_p=43$} & \multicolumn{3}{c}{$\mathit{top}_p=35$} & \multicolumn{3}{c}{$\mathit{top}_p=25$} & \multicolumn{3}{c}{$\mathit{top}_p=15$} \\
        \cmidrule(lr){2-4} \cmidrule(lr){5-7} \cmidrule(lr){8-10} \cmidrule(lr){11-13}\\
        \addlinespace[-8pt]
        {} & $1^\mathit{st}$ & $2^\mathit{nd}$ & $3^\mathit{rd}$ & $1^\mathit{st}$ & $2^\mathit{nd}$ & $3^\mathit{rd}$ & $1^\mathit{st}$ & $2^\mathit{nd}$ & $3^\mathit{rd}$ & $1^\mathit{st}$ & $2^\mathit{nd}$ & $3^\mathit{rd}$ \\
        \midrule
        $\mathcal{N}_\mathit{SN}$ & $1$ & $0$ & $-$ & $1$ & $0$ & $-$ & $0$ & $-$ & $-$ & $0$ & $-$ & $-$ \\
        $\mathcal{N}_\mathit{MN}$ & $1$ & $2$ & $1$ & $2$ & $0$ & $-$ & $1$ & $0$ & $-$ & $0$ & $-$ & $-$ \\
        $\mathcal{N}_\mathit{LN}$ & $2$ & $0$ & $-$ & $1$ & $2$ & $0$ & $0$ & $-$ & $-$ & $0$ & $-$ & $-$ \\
        \midrule
        $\mathcal{N}_\mathit{SN}^\mathit{troj}$ & $3$ & $0$ & $-$ & $3$ & $0$ & $-$ & $1$ & $1$ & $0$ & $0$ & $-$ & $-$ \\
        $\mathcal{N}_\mathit{MN}^\mathit{troj}$ & $2$ & $0$ & $-$ & $2$ & $0$ & $-$ & $1$ & $2$ & $0$ & $1$ & $0$ & $-$ \\
        $\mathcal{N}_\mathit{LN}^\mathit{troj}$ & $1$ & $0$ & $-$ & $1$ & $0$ & $-$ & $0$ & $-$ & $-$ & $0$ & $-$ & $-$ \\
        \bottomrule
    \end{tabular}
    \label{Table: Multiple Mitigation}
\end{table}

\begin{table}[htbp]
    \setlength{\tabcolsep}{4.pt} 
    \centering
    \caption{Number of trojan neurons and their ASR after neuron pruning on trojan models}
    \begin{tabular}{lcccccccc}
        \toprule
        \multirow{2}{*}[-0.25em]{Model} & \multicolumn{4}{c}{$\#$ of trojan neurons} & \multicolumn{4}{c}{Attack success rate}\\
        \multirow{2}{*}[-0.25em]{} & \multicolumn{4}{c}{at different pruning rates} & \multicolumn{4}{c}{at different pruning rates}\\
        \cmidrule(lr){2-5} \cmidrule(lr){6-9} \\
        \addlinespace[-8pt]
        {} & Before & $0.4$ & $0.5$ & $0.6$  & Before & $0.4$ & $0.5$ & $0.6$ \\
        \midrule
        $\mathcal{N}_\mathit{SN}^\mathit{troj}$ & $4$ & $4$ & $4$ & $4$ & $97.69\%$ & $97.68\%$ & $97.68\%$ & $97.68\%$ \\
        $\mathcal{N}_\mathit{MN}^\mathit{troj}$ & $3$ & $3$ & $3$ & $3$ & $93.00\%$ & $80.24\%$ & $97.83\%$ & $97.83\%$ \\
        $\mathcal{N}_\mathit{LN}^\mathit{troj}$ & $2$ & $1$ & $1$ & $1$ & $80.72\%$ & $53.07\%$ & $53.07\%$ & $53.07\%$ \\
        \bottomrule
    \end{tabular}
    \vspace{-0.50cm}
    \label{Table: Pruning Iterations}
\end{table}

As an additional experiment, we evaluate the effect of \emph{neuron weight pruning}~\cite{wang2019neural} on the trained models.
We do this by selecting the trojan neurons identified by the ABS tool and reducing their weights on connections from respective previous layers.
This way, we hope to reduce the information flow through these trojan neurons by a certain percentage which we call it as \emph{pruning rate}.
Pruning rate takes values between $0$ (no change in the weights) and $1$ (all the weights set to $0.0$). 
The results depicted in Table~\ref{Table: Pruning Iterations} show that the weight pruning do not reduce the number of trojan neurons.
This may be because unlike~\cite{wang2019neural}, we use ABS to identify trojan neurons and the number of trojan neurons we obtain is very low for this analysis.
It is interesting to exploit better pruning technique which could lead to a better mitigation performance. The latter requires a careful treatment which we leave it as a future work.

We directly profit from the advantages of using the ABS tool instead of Neural Cleanse which are discussed in~\cite{liu2019abs}.
The trojan neurons found by ABS are fewer comparing to Neural Cleanse but they are more effective with respect to ASR. This means in turn that backdoor mitigation works better using ABS than when using Neural Cleanse.
More important however is that our retraining method works better. Our results demosntrates that, in contrast to Neural Cleanse, strategically retraining the model using masked images from $\mathit{top}_p$ classes can remove all identified backdoors or biases in the model.
% With our results, we show that, in contrast to Neural Cleanse, strategically retraining the model using masked images from $\mathit{top}_p$ classes, can remove all identified backdoors/biases in the model.
Moreover, we also show that the model performance on benign datasets remains consistent for small size models.
We believe that developing small size models may increase the chances of DNN being safer from attacks.

\section{Conclusion}
In this paper, we have addressed the problem of backdoor mitigation in classification models.
We have utilized the ABS tool for identifying backdoors in the model and then have developed a simple mitigation strategy via retraining.
Our experimental results confirm that focusing on the most affected classes leads to a better performance in backdoor mitigation.

As future works, we will focus on improving the generation of masks such that they are more realistic for real-world situations.
Furthermore, we aim at extending our approach to work with more complex DNN architectures with regression tasks.
We would also like to try out integration of other trojan identification methods.

\bibliographystyle{splncs04}
\bibliography{Bibliography}

\end{document}